\documentclass[aps,pre,preprint,onecolumn,showpacs,amsmath,amssymb]{revtex4-1}

\usepackage{graphicx}
\usepackage{bm}
\usepackage{textgreek}

\begin{document}

\title{Tubulation and dispersion of oil by bacterial growth on droplets}

\author{Vincent Hickl$^{a}$}

\author{Gabriel Juarez$^{b}$}
\thanks{Email address}\email{gjuarez@illinois.edu}

\affiliation{$^{a}$Department of Physics, University of Illinois at Urbana-Champaign, Urbana, Illinois 61801, USA}

\affiliation{$^{b}$Department of Mechanical Science and Engineering, University of 
Illinois at Urbana-Champaign, Urbana, Illinois 61801, USA}

\date{\today}


\begin{abstract}


Bacteria on surfaces exhibit collective behaviors, such as active turbulence and active stresses, which result from their motility, growth, and interactions with their local surroundings. However, interfacial deformations on soft surfaces and liquid interfaces caused by active growth, particularly over long time scales, are not well understood. Here, we describe experimental observations on the emergence of tubular structures arising from the growth of rod-shaped bacteria at the interface of oil droplets in water. Using microfluidics and timelapse microscopy, the dimensions and extension rates of individual tubular structures as well as bulk bio-aggregate formation are quantified for hundreds of droplets over 72 hours. Tubular structures are comparable in length to the initial droplet radius and are composed of an outer shell of bacteria that stabilize an inner filament of oil. The oil filament breaks up into smaller microdroplets dispersed within the bacterial shell. This work provides insight into active stresses at deformable interfaces and improves our understanding of microbial oil biodegradation and its potential influence on the transport of droplets in the ocean water column.

\end{abstract}

\pacs{XX.xx.xx}

\maketitle

\section{Introduction}

Active matter refers to systems that exchange energy with their surroundings, leading to behaviors that are unlikely at thermal equilibrium. Active systems consisting of simple building blocks can give rise to complex and emergent phenomena. Examples range across many scales, from actin microfilaments and clusters of cells \cite{Doostmohammadi2018,Copenhagen2016} to flocks of birds \cite{Bialek2014} and schools of fish \cite{Herbert-Read2011}. The study of active materials shows promise for uncovering new physics, with potential applications in a variety of disciplines, including statistical and biological physics \cite{Bechinger2016,Ramaswamy2010}, soft matter \cite{Doostmohammadi2018,Marchetti2013}, biomedicine \cite{Wang2012}, materials science \cite{Needleman2017}, and traffic engineering \cite{Helbing2001}.

Active particles confined to interfaces form a rich subset of active systems of particular interest to physical and biological research communities. Both living and non-living interfacial systems can exhibit complex behaviors far from thermal equilibrium including collective motility, swarming, and active turbulence \cite{Copenhagen2016,Metselaar2019,Guillamat2017}. These emergent behaviors are believed to play an important role in biological phenomena such as cellular transport \cite{Sanchez2012,Woodhouse2012}, self-assembly of microtubules and motor proteins \cite{Kruse2004,Surrey2001}, clusters of sperm or epithelial cells \cite{Creppy2015,Blanch-Mercader2018}, and bacterial or algal colonies \cite{Lauga2006,Berke2008}. 

Because of their simplicity, single-celled bacteria are an especially useful model organism to explore how complex behavior emerges in collections of simple components \cite{Aranson2022}. Studies have explored the ways in which monolayers of motile bacteria give rise to active behavior \cite{Cisneros2007}. Other researchers have created artificial micro- or nanoswimmers to mimic the self-propulsion observed in living cells \cite{Gompper2020,Tsang2020}. Rod-shaped bacteria in particular are known for their tendency to align and self-organize at interfaces \cite{Doostmohammadi2018}. The orientational ordering of such bacteria can greatly increase the magnitude of the stresses exerted on the substrate by either motility or growth of the cells. However, self-organization of bacterial colonies has mostly been studied on rigid interfaces, where interfacial deformations can be neglected. In many systems this assumption does not always hold, and possible re-organization of the active material's environment must be considered \cite{Bechinger2016}. 

Other active systems have been studied for the interfacial deformations that can result from the collective motion of active particles. For example, self-propelled particles enclosed within giant vesicles can induce a variety of non-equilibrium shapes in the membrane \cite{Vutukuri2020,Alert2022}. A common deformation mode in active systems is tubulation: the emergence of high-aspect-ratio tubular structures from an interface. This process has been studied extensively of liposomes or vesicles, where a phospholipid bilayer folds onto itself to form a protrusion from the cell \cite{Lamaziere2007, LienRoux2002, Osawa2009}. Tubulation can also be driven by a layer of filaments, such as FtsZ or microtubules, that coat the surface of the vesicle \cite{Keber2014,Kumar2019}. Recently, generic models have been developed to describe how active nematics, such as collections of rod-shaped particles, produce tube-like deformation on deformable interfaces \cite{Metselaar2019,Hoffmann2022}. To verify the generality  of these results, it is important to determine experimentally whether other model active nematics such as rod-shaped bacteria can produce similar morphologies. 

The study of bacteria at oil-water interfaces is important because of its relevance to oil spill remediation research. After a deep-sea oil spill, oil-degrading bacteria such as \emph{Alcanivorax borkumensis} \cite{Yakimov1998} play a crucial role in removing oil from marine ecosystems \cite{Head2006,Passow2016,Passow2021}. The process of biodegradation has primarily been studied using\textit{ in situ} samples and batch reactors \cite{Kleindienst2015,Brakstad2014,Prince2014}, which cannot reveal the underlying physics at the oil-water interface. Many open questions remain, particularly regarding how microscopic droplets of oil in sub-surface plumes can be biodegraded \cite{Hazen2010,McFarlin2014}. 

The effect of droplet size on degradation rates (including the effect of surfactants and chemical dispersants) is especially contentious \cite{Kleindienst2015,Prince2015}. Recent studies have examined how oil-degrading bacteria attach to oil droplets and form aggregates such as biofilms at the interface \cite{Bookstaver2015,Abbasi2018,Omarova2019}. However, none have considered the complex deformations to the oil-water interface that can arise as a result of cell growth. Studies of individual droplets under environmentally relevant conditions are needed to better understand the fate of spilled oil \cite{Lee2013,White2019}. In particular, direct observation of individual droplets with high spatial and temporal resolution can shed light on the biomechanics of oil biodegradation.

The self-organization and collective behavior of growing bacteria at liquid interfaces has recently been shown to lead to the colonization and buckling of oil droplets \cite{Fernandez2022, Hickl2022}. Here, we describe the extreme deformations of the oil-water interface and dispersion of oil caused by the growth of bacteria on spherical droplets. Using microfluidics and timelapse microscopy, hundreds of oil droplets of sizes $5$ \textmu m $\leq R\leq 100 $ \textmu m are observed simultaneously and continuously over 72 hours. Confined cell growth leads to tubulation at the interface. We describe the underlying structure and composition of the tubes, which can form a dense bio-aggregate in which oil is dispersed as smaller droplets. Image analysis is used to measure the tube and bio-aggregate extension rate, bio-aggregate size, and size distribution of micro-dispersed droplets. These phenomena demonstrate how the growth of simple, non-motile active particles on an interface can lead to the emergence of complex interfacial deformations. The ability to visualize these processes has important benefits for the study of bacterial biodegradation of oil and, more generally, the interactions between fluid interfaces and active matter.


\section{Experimental methods}

The species used in experiments was \textit{Alcanivorax borkumensis}, a known anaerobic oil-degrading bacterium. Cells are rod-shaped which average length $2.7$ \textmu m and width $0.7$ \textmu m. Cells were grown in a culture medium consisting of 37.4 g/L 2216 marine broth (BD Difco) and 10 g/L sodium pyruvate in an orbital shaker at 180 rpm and $30 ^{\circ}$C. At the start of each experiment, a cell culture in the late exponential phase is diluted in culture medium to an optical density $OD=0.18$, measured using a Biowave CO8000 cell density meter. Then, the suspension was further diluted $50\times$ in artificial seawater (ASW). This process ensures that the initial cell concentration in all experiments is the same. The mean buckling time of the cells in culture medium was measured to be $1.6$ hours.

A custom microfluidic chamber was constructed to visualize the time evolution of oil droplets colonized by bacteria \cite{Hickl2022}. To create droplets on a glass microscope slide, a thin film of $10$ \textmu L of either unweathered MC252 crude oil or mineral oil was formed. MC252 is a light sweet crude oil with a density of $0.83$ g/mL, an interfacial tension of $20$ mN/m, and a viscosity of $3.9$ mPa s at $32^{\circ}$C \cite{Daling2014}. Then, $450$ \textmu L of the dilute \textit{A. borkumensis} suspension was dripped on top of the oil. This aqueous solution initiated the formation of spherical caps pinned to the glass slide. The droplet size range was $1$ \textmu m $< R < 200$ \textmu m. The caps contained $93\pm3\%$ of the volume of a sphere with the same diameter, on average. Then, inlets and outlets made of polyethylene tubing were attached to both ends of the device. Lastly, the chamber was sealed using silicone rubber sheets (McMaster-Carr) and epoxy. The chamber dimensions ($L\times W\times H$) were $3\times 1\times0.15\text{ cm}$. 

To maintain cultures within the chamber for 72 hours, a syringe pump was used to continuously inject a solution of diluted culture medium and ASW (1:9). In the experiments where bacteria cells were stained, a solution of 5 mM SYTO 9 nucleic acid stain (Sigma Aldrich) was added to the solution of diluted culture medium and ASW. These solutions were injected into the chamber at a flow rate of $0.01$ mL/min, generating an average flow speed of $11$ \textmu m/s and a Reynolds number of $\textrm{Re}= \rho U D/ \mu = 0.03$. Here, $\rho$ is the fluid density, $U$ is the mean flow speed, $D$ is the equivalent diameter of the channel, and $\mu$ is the fluid viscosity. Experiments were performed at a room temperature of $20 ^\circ$C.

Using timelapse phase contrast and epi-fluorescence microscopy, images of over $100$ droplets were taken every 10 minutes in each experiment. Image capture of each droplet was performed at 40X magnification using a Nikon Ti Eclipse equipped with an automated motor-controlled stage and an Andor Zyla 4.2 sCMOS camera.

\section{Results}

\subsection{Oil droplet evolution}

\begin{figure*}
\centering
    \includegraphics[width=\linewidth]{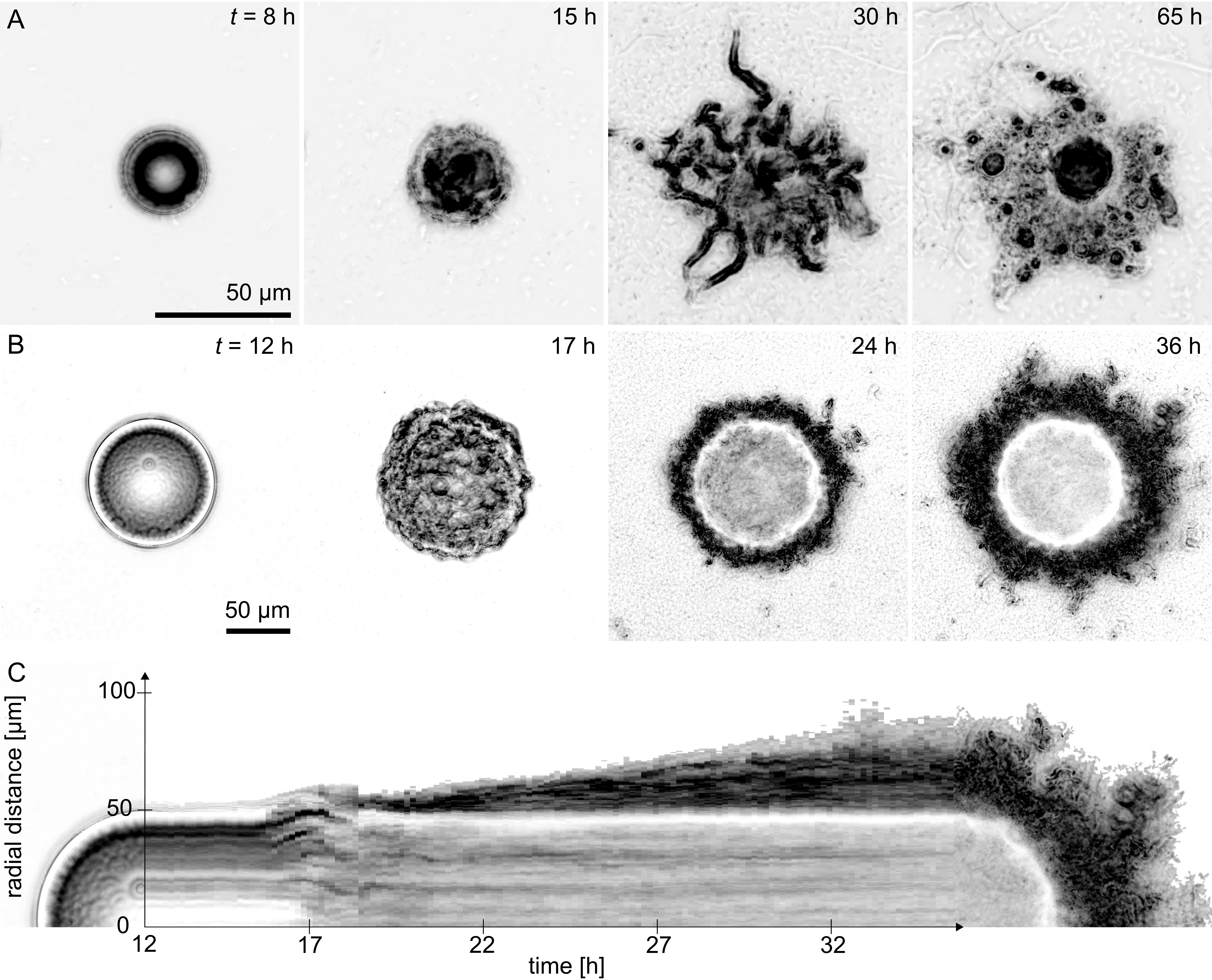}
\caption{
(A) The evolution of an oil droplet ($R_0=16$ \textmu m) as it is deformed by the confined growth of bacteria.
Colonization occurs ($t\leq 10$ h) as bacteria grow and divide at the droplet surface. 
Buckling occurs ($t\approx 15$ h) after the droplet surface is completely saturated by bacteria. 
Tubulation occurs ($15\leq t\leq35$ h) as the oil-water interface continues to deform because of the growth of bacteria. 
Secondary droplet formation and dispersion of oil occurs at later times ($35\leq t\leq72$ h).
See supplementary material for example timelapse videos. 
(B) The evolution of a larger droplet ($R_0=51$ \textmu m) is qualitatively similar. However, tubes form a dense bio-aggregate around the droplet.
(C) The evolution of the droplet in (B) plotted as a spacetime plot, showing the expansion ($18\leq t\leq 36$ h) of the bio-aggregate from the interface. 
}
    \label{fig:panel1}
\end{figure*}

Prolonged growth of bacteria on oil droplets leads to extreme deformations of the interface. The mechanism driving this process is the competition between the effects of cell growth and the surface tension of the oil-water interface. Images of the time evolution of two droplets of radii $R_0=16$ \textmu m and $R_0=51$ \textmu m are shown in Fig. \ref{fig:panel1}A and 1B, respectively (see also Supplementary videos 1 and 2). Buckling of the interface occurs shortly after bacteria fully cover the droplet surface, shown in Fig. \ref{fig:panel1}A, $t=15$ h and Fig. \ref{fig:panel1}B, $t=17$ h. Then, tube-like structures emerge from the interface, extending outward from the droplet surface. For droplets with initial radius $R\lesssim15$ \textmu m, individual tubes remain distinct, as shown in Fig. \ref{fig:panel1}A, $t=30$ h. By contrast, for droplets with $R\gtrsim15$ \textmu m, tubes become densely packed, forming a continuous bio-aggregate around a droplet, shown in Fig. \ref{fig:panel1}B, $t=36$ h . These deformations are qualitatively the same for both crude oil and mineral oil.

To visualize the rich dynamics of the system, including the expansion of the bio-aggregate, vertical slices were taken through the center of each image in a timelapse of a droplet's evolution. These slices were then arranged next to one another to form a spacetime plot showing bio-aggregate expansion over time. For example, the spacetime plot of the expansion of the bio-aggregate around a droplet with $R_0=51$ \textmu m is shown in in Fig. \ref{fig:panel1}C. Over 1--2 days after tubes and bio-aggregates stop growing, oil within the tubes breaks up and coalesces into spherical droplets, shown in Fig. \ref{fig:panel1}A, $t=65$ h. 


\subsection{Bio-aggregate composition}

To determine the composition of the bio-aggregate, fluorescence microscopy was used to visualize the distribution of crude oil after droplets were deformed, since the crude oil used is autofluorescent in the visible spectrum. In other experiments, cells were dyed with SYTO 9 nucleic acid stain to visualize the distribution of bacteria around deformed droplets. To visualize only the bacteria, non-fluorescent mineral oil was used in experiments with SYTO 9. Phase contrast microscopy was also used to observe all parts of the system simultaneously regardless of the type of oil. Thus, the distribution of both oil and bacteria around deformed droplets was determined. 

\begin{figure*}
    \centering
    \includegraphics[width=\linewidth]{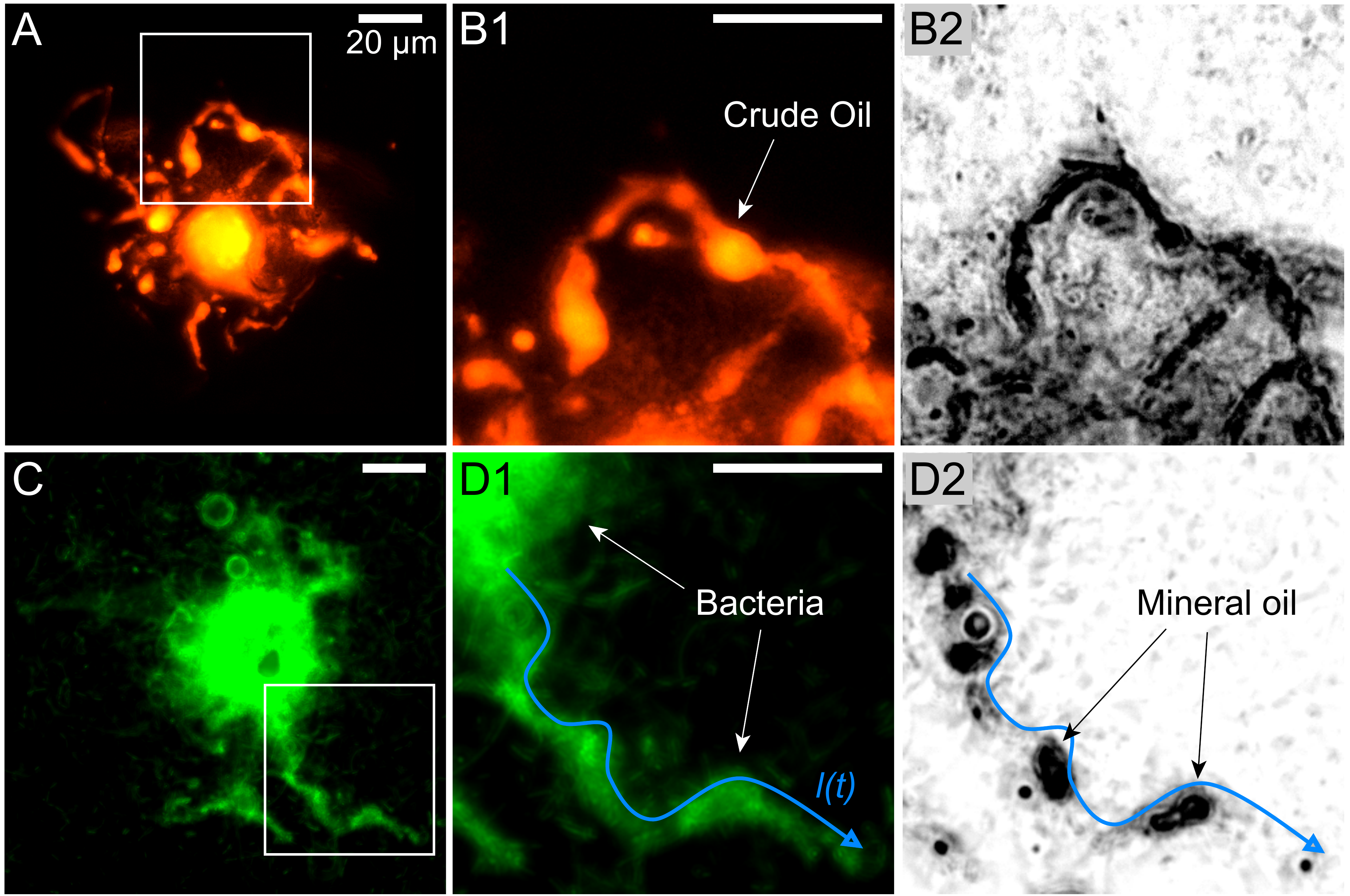}
\caption{Bio-aggregate structure and oil dispersion. (A) Epifluorescence microscopy shows crude oil (orange false color) at the center of a deformed droplet ($R_0=15$ \textmu m), and contained within the tube structures surrounding the droplet. Magnified views of a tube containing crude oil with (B1) epifluorescence and (B2) phase contrast microscopy, show oil within the tube and the surrounding bio-aggregate, respectively. (C) A mineral oil droplet ($R_0=18$ \textmu m) colonized by bacteria stained with SYTO 9 appears bright at the center and along the tube structures (green false color) surrounding the droplet. 
Magnified views of a tube structure with (D1) epifluorescence (D2) phase contrast microscopy show fluorescent bacteria along the tube and the mineral oil droplets contained within the same tube, respectively. The contour length of the tube $l(t)=74$ \textmu m is shown in blue in panels D1 and D2.
All scale bars are $20$ \textmu m.} 
    \label{fig:panel2}
\end{figure*}

The tubes protruding from the oil-water interface contain both oil and bacteria. An example of a deformed crude oil droplet of $R_0=15$ \textmu m after 50 hours is shown in Fig.~\ref{fig:panel2}A. The orange (false color) corresponds to crude oil when visualized using fluorescence microscopy. Oil is present throughout the tube of length 60 \textmu m, which extends from the droplet surface in Fig. \ref{fig:panel2}B1. When visualized using phase contrast, the oil (black) can be seen in sharp contrast against the surrounding bacteria (grey) and device background (white), as in Fig. \ref{fig:panel2}B2. A mineral oil droplet of initial radius $R_0=18$ \textmu m after 48 hours is shown in Fig. \ref{fig:panel2}C. Bacteria can be seen in green (false color) around both the original spherical droplet and a tube of length $74$ \textmu m protruding from the surface in Fig. \ref{fig:panel2}D1. As before, phase contrast images show the dark crude oil present within the tube shown in Fig. \ref{fig:panel2}D2. Together, these images show that the tubular structures in our experiments are tubes of oil surrounded and stabilized by a shell of bacteria.

\subsection{Tube extension}

Having determined the composition of the tubes, the dynamics of their extension from the droplet surface over time are now described. The contour length $l(t)$ of a tube is represented by the blue line in Fig. \ref{fig:panel2}D, where $l(t)=74$ \textmu m. The contour length was measured in successive frames of the timelapse for tubes that remained in focus throughout the experiment. The extension of ten representative tubes from different droplets over 4-7 hours is shown in Fig. \ref{fig:panel3}A, inset. Here $t=0$ corresponds to the time when each tube begins to extend from the droplet surface. For the first few hours, $l(t)$ appears to increase exponentially with time. Final tube lengths are in the range $10$ \textmu m $\leq l\leq75$ \textmu m with a mean of $l=33\pm1$ \textmu m based on 125 different tubes. 



\subsection{Bio-aggregate expansion}

\begin{figure}
\centering
    \includegraphics[width=\linewidth]{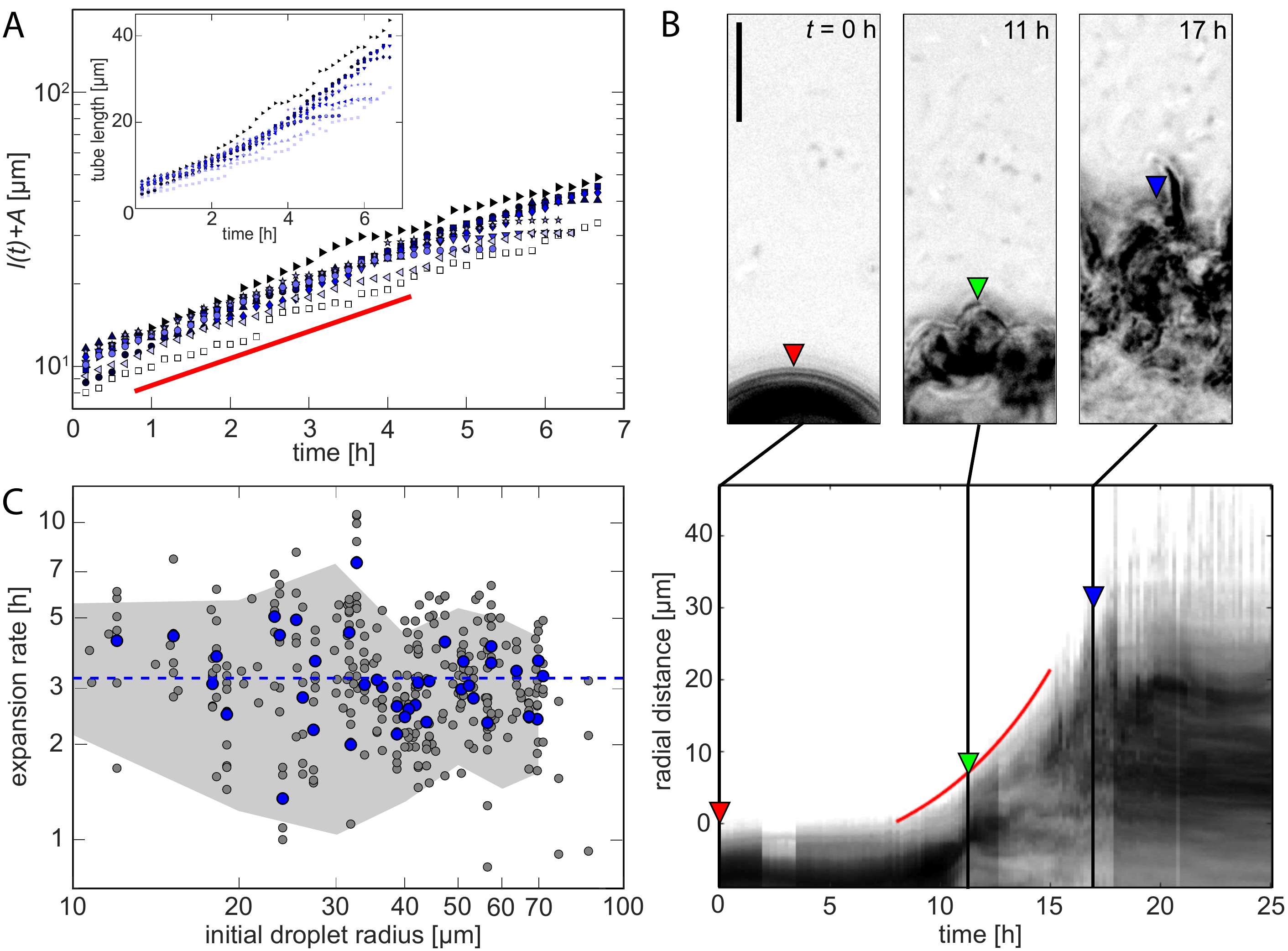}
\caption{(A) Inset: Tube length for 10 representative tubes extending over time from different droplets. Main panel: fitting of experimental data according to mechanistic model of tube extension. The quantity $l(t)+A$ is plotted over time on a semilog scale. Solid red line shows mean exponential fit: $l(t)+A\propto e^{\ln{(2)}t/3.1\text{h}}$. (B) Measurements of bio-aggregate size at three different times (top), and corresponding spacetime plot (bottom) showing aggregate expansion from the surface of a crude oil droplet with $R_0 = 18$ \textmu m. Expansion rates are measured by fitting an exponential curve (red line) to the boundary of the aggregate during the initial outward increase in size. Scale bar is 10 \textmu m. (C) The aggregate expansion rates for a representative sample of crude oil droplets of varying initial radius. Gray points represent individual expansion rate measurements (as shown in panel (B)) at different positions along a droplet's surface. Blue points represent the average expansion rate ($N=41$) for $5-10$ positions around a single droplet. The dashed line shows the mean expansion rate for all droplets. The shaded area shows the mean $\pm1.5$ standard deviations as a function of droplet radius. Above $R_0=70$ \textmu m, insufficient data were obtained to calculate these statistics. The expansion rate does not depend on the initial droplet size ($p>0.05$).}
    \label{fig:panel3}
\end{figure}

Next, the expansion over time of bio-aggregates, which consist of many densely packed tubes of oil and bacteria, is described. The bio-aggregate expansion from a point on a droplet with $R_0=18$ \textmu m at $t=0$, $11$, and $17$ h after buckling is shown in Fig. \ref{fig:panel3}B. A spacetime plot was created by taking a vertical slice of each frame in the timelapse and arranging them next to one another, as in the bottom half of Fig. \ref{fig:panel3}B. This plot shows the expansion of the bio-aggregate. The colored arrows show the edge of the bio-aggregate, which was used to measure the radial extent of the aggregate from the droplet surface in each frame of the timelapse. As with individual tube length, the initial aggregate expansion over time is exponential. An exponential curve fitted to the data is shown Fig. \ref{fig:panel3}B, red line. Here the doubling time of the aggregate is $3.78$ hours. 

The expansion rate is, on average, independent of initial droplet radius. The expansion rate was measured at 5--10 different points along the perimeter of 42 different droplets with $10\text{ \textmu m}\leq R_0\leq100$ \textmu m, giving a total of 323 measurements of the expansion rate, shown in Fig. \ref{fig:panel3}C, grey points. Averaging the measurements from each droplet gives the mean expansion rate for that droplet, shown in Fig. \ref{fig:panel3}C, blue points. Mean expansion rates range from 1.3 h to 7.5 h (including outliers), with 88\% of measurements ranging from 2 h and 5 h. For all droplets, the mean is $3.2\pm1.1$ h, as shown in Fig. \ref{fig:panel3}C, blue dashed line. The expansion rate does not depend on the initial droplet size ($p>0.05$). 

\begin{figure*}
\centering
    \includegraphics[width=\linewidth]{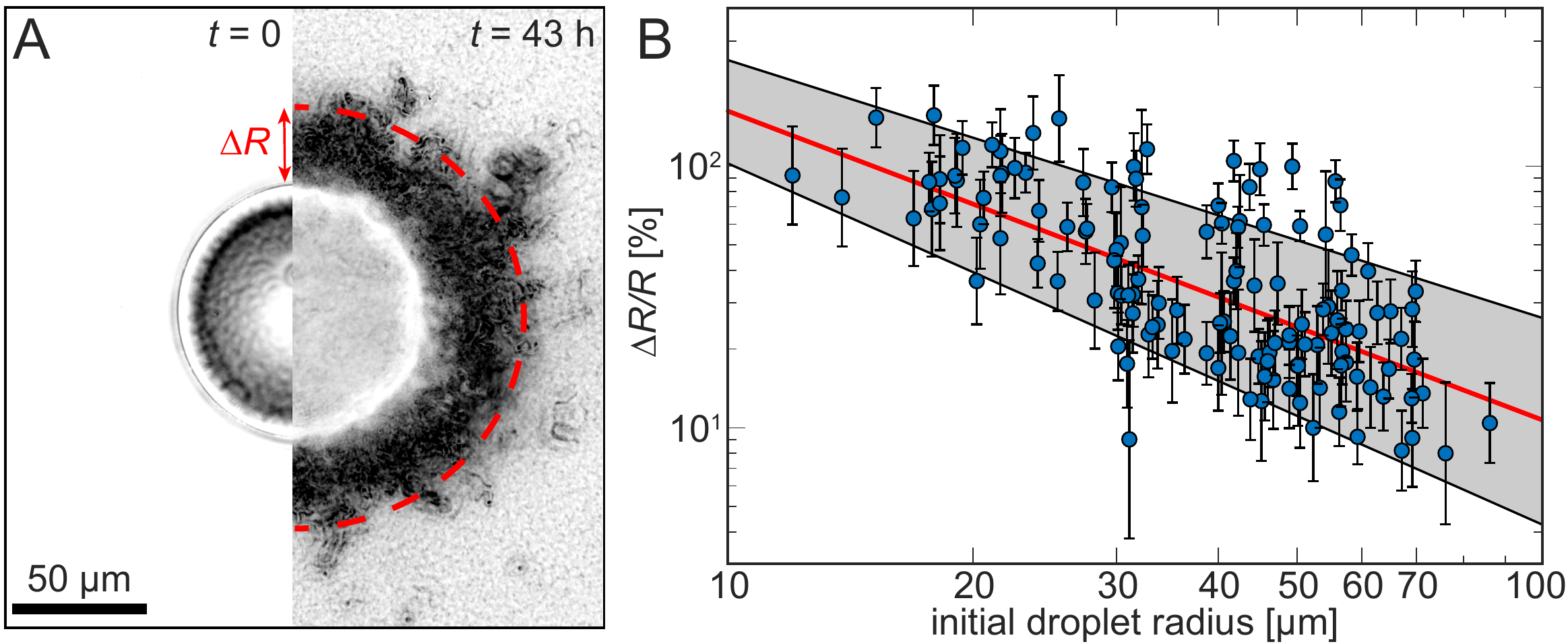}
\caption{
Bio-aggregate formation and increase in droplet size caused by tubulation.
(A) Comparison of a crude oil droplet before ($t=0$ h) and after ($t=43$ h) it has been colonized and deformed by bacteria. Here, $R_0=51$ \textmu m and the emergent tubes have increased the final bio-aggregate radius by an amount $\Delta R=30$ \textmu m.
(B) The final bio-aggregate radius with respect to the initial droplet radius for $131$ droplets. Each point represents the average value of $\Delta R$ for a single droplet, and the error bar shows the variation along the droplet perimeter. The best power law fit to the data (red line) corresponds to a slope of $-1.2$ and the shaded area indicates the 95\% confidence interval for the fit. 
}
    \label{fig:panel4}
\end{figure*}

The emergence of a bio-aggregate caused by interfacial growth of bacteria significantly increases the effective size of oil droplets. A side-by-side comparison of a droplet of crude oil with $R_0=51$ \textmu m before ($t=0$) and after ($t=43$ h) it is colonized and deformed by bacteria is shown in Fig. \ref{fig:panel4}A. The bio-aggregate size $\Delta R$ is defined as the distance between initial surface of the droplet at $t=0$ and the edge of the aggregate at the end of the experiment. This distance is measured radially at many points along the droplet's perimeter. The mean aggregate size for this droplet is $\Delta R=30$ \textmu m, or $60\%$ of the initial droplet radius, as shown in Fig. \ref{fig:panel4}A, red line. In fact, $\Delta R$ is approximately equal to the mean final tube length of $l=33$ \textmu m, shown in \ref{fig:panel3}A, inset.

The relative increase in droplet size due to aggregate expansion is larger for small droplets. The quantity $\Delta R/R$ for 131 droplets is shown in Fig. \ref{fig:panel4}B. Error bars represent the variation in bio-aggregate size along the droplets' perimeters. The change in size varies from $8\%$ to $156\%$, with a mean value of $46\%$. The value of $\Delta R$ itself ranges from $3$ \textmu m to $49$ \textmu m. Power law fitting (red line) shows that $\Delta R/R \propto R^{-1.20\pm0.23}$, suggesting that $\Delta R$ is, on average, independent of the initial droplet radius $R$, and that tube length is independent of droplet size. Consequently, the relative increase in total size tends to be greater for smaller droplets.



\subsection{Secondary droplet formation}

\begin{figure*}
    \centering
    \includegraphics[width=\linewidth]{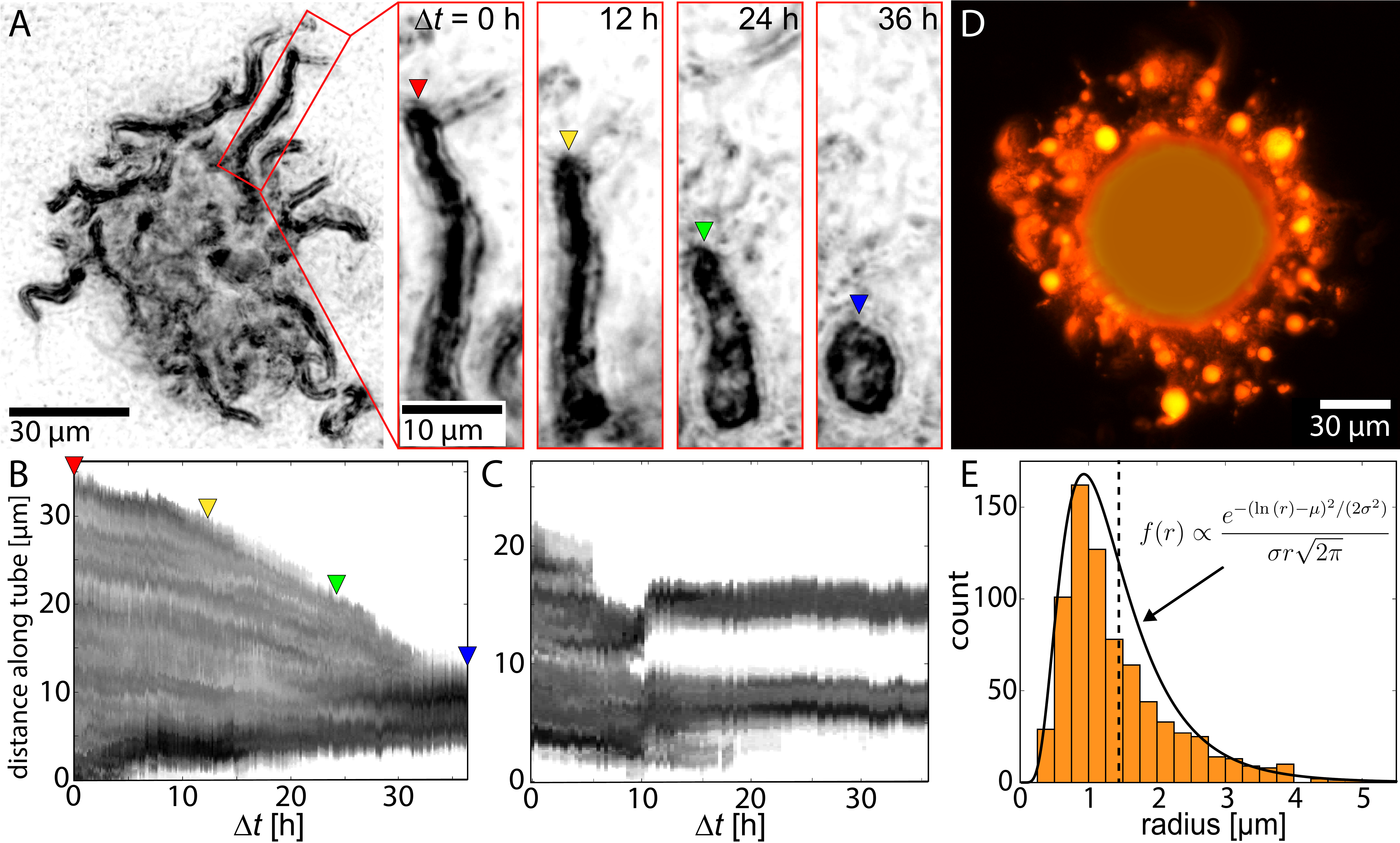}
\caption{
Secondary droplet formation in the bio-aggregate.
(A) Droplet with initial radius $R_0=17$ \textmu m after tube formation. Four snapshots ($\Delta t=0$, 12, 24, and 36 h) of secondary droplet formation inside a single tube are shown. Colored arrows show the furthest extent of the oil within the tube in each frame.
(B) Movement of oil within the tube above, plotted as a kymograph, as it coalesces into a single spherical droplet of $R=4.9$ \textmu m. The data points from the inset of panel A are indicated by the corresponding arrows.
(C) Kymograph showing the movement of the oil in a different tube. Here, the oil breaks up and coalesces into two secondary droplets of radii $r=1.9$ \textmu m and $r=2.1$ \textmu m, respectively.
(D) Secondary droplets embedded in a bio-aggregate surrounding the primary droplet visualized using fluorescent microscopy. 
(E) Size distribution of secondary droplet radii measured for $753$ secondary droplets at the end of the 72 hours of droplet evolution. The lognormal fit to the data with $\mu=0.21$ and $\sigma=0.53$ is shown (black line). The dashed black line shows the mean secondary droplet radius $r=1.48$ \textmu m.
}
    \label{fig:panel5}
\end{figure*}

Finally, the changes in oil morphology within tubes are examined, which occur over 1--2 days after tubes and bio-aggregates form. The oil within each cylindrical tube coalesces into one or more spheres, or secondary droplets. An extremely deformed droplet with $R_0=17$ \textmu m after 21 hours is shown in Fig. \ref{fig:panel5}A. Insets show the evolution of the oil (dark) within a tube of length $l=33$~\textmu m over 36~hours. Initially the oil fills the entire length of the tube, but then shrinks toward the base of the tube and coalesces into a spherical droplet of radius $r=4.9$ \textmu m at 36 hours (see Supplementary video 3). A spacetime plot qualitatively describes the evolution of oil coalescing within this tube, shown in Fig. \ref{fig:panel5}B. Colored arrows show the edge of the oil within the tube at different times. The oil in different tubes can break up into between 1 and 7 smaller droplets. A spacetime plot of oil breaking up and coalescing into 2 distinct droplets of radii $r=1.9$ \textmu m and $r=2.1$ \textmu m, respectively, is shown in \ref{fig:panel5}C (see also Supplementary video 4).  

After the system reaches a steady state, fluorescence microscopy was used to observe the secondary droplets embedded in the bio-aggregate, as shown in Fig. \ref{fig:panel5}D. Using a \textit{z}-stack with height increments of 2 \textmu m, all secondary droplets within the bio-aggregate were identified and their sizes measured. A histogram of the radii of secondary droplets ($N=753$) is shown in Fig. \ref{fig:panel5}E. These data were taken from tubes and bio-aggregates surrounding many different droplets with $10 \text{ \textmu m}\leq R_0\leq 100 \text{ \textmu m}$. Secondary droplet radii are in the range $0.25\text{ \textmu m}\leq r\leq 6 \text{ \textmu m}$ with a mean of $\Bar{r}=1.48 \pm 0.03 \  $\textmu m --- more than an order of magnitude smaller than the original droplet sizes. The secondary droplets' size appears to follow a lognormal distribution:
\begin{equation}
    f(r)\propto\frac{e^{-(\ln{(r)}-\mu)^2/(2\sigma^2)}}{\sigma r\sqrt{2\pi}}
\end{equation}
with parameters $\mu=0.21\pm0.02$ and $\sigma=0.53\pm0.01$, shown in Fig. \ref{fig:panel5}E, black line. 
    

\subsection{Model of tube extension}

Since tube and bio-aggregate expansion are exponential, and bacteria appear to cover the oil-water interface throughout the experiment, we suggest that the emergence of these complex structures is driven solely by the growth and division of cells confined to the interface. Here we present a mechanistic model of tube formation to quantify the relationship between cell growth and tubulation. 

We assume that tubes begin to extend from a droplet when bacteria have formed a complete monolayer of cells at the interface. To continue to grow at the interface, cells must deform the oil droplet to increase the surface area. Assuming that the density of cells at the interface remains constant, and that cells remain as a monolayer, the interfacial area must increase proportionally to the number of cells:
\begin{equation}
    A(t)=4\pi R_0^2 \ e^{\ln(2)t/T_d}\text{,}
\end{equation}
where $T_d$ is the cell doubling time, and $R_0$ is the initial radius of the droplet. At $t=0$, tubes of oil begin to emerge from the interface. We assume that the total volume of oil remains constant (i.e., the oil is incompressible and loss of volume from degradation is negligible):
\begin{equation}
    V(t)=V(0)=\frac{4}{3}\pi R_0^3
\end{equation}
Over time, the total surface area and volume of the oil-water interface is given by
\begin{equation}
    A(t)=4\pi R(t)^2 + 2\pi r \ l(t)N
\end{equation}
\begin{equation}
    V(t)=\frac{4}{3}\pi R(t)^3 + N\pi r^2 \ l(t)\text{,}
\end{equation}
where $N$ is the number of tubes, $r$ is the tube radius, and $R(t)$ and $l(t)$ are the time-varying droplet radius and tube contour length, respectively. Combining Eqns. (3), (4) and (5) gives
\begin{equation}
    A(t)=4\pi R_0^2 \left[1-\frac{3N}{4}\frac{l(t)r^2}{R_0^3}\right] ^{2/3} + 2\pi rl(t)N\text{.}
\end{equation}
For values of the parameters pertaining to this system, the second term in brackets is much less than unity. That is, up to a constant, the volume of oil in the tubes is much less than the total volume, which is in qualitative agreement with observations. So, to first order in $l(t)r^2/R_0^3$, we find that:
\begin{equation}
    A(t)\approx 4\pi R_0^2 + 2\pi Nr\left(1-\frac{r}{R_0}\right)l(t)\text{.}
\end{equation}
Finally, we substitute Eqn. (2) and $N=4\pi R_0/d^2$, where $d$ is the mean separation of tubes at the interface, to find:
\begin{equation}
    l(t)=\frac{d^2}{2\pi r\left(1+\frac{r}{R_0}\right)}\left(e^{\frac{\ln(2)t}{T_d}}-1\right)\text{.}
\end{equation}
Thus, the model predicts that, because of the exponential growth of bacteria at the interface, the tube length plus a constant increases exponentially with time:
\begin{equation}
    l(t)+A=Ae^{\frac{\ln(2)t}{T_d}}\text{,}
\end{equation}
where
\begin{equation}
    A=\frac{d^2}{2\pi r\left(1+\frac{r}{R_0}\right)}.
\end{equation}

To test the predictions of Eqn. (7), $l(t)+A$ is plotted on a semilog scale, shown in Fig. \ref{fig:panel3}A. The following experimental values were used to compute $A$: $d=5.9$ \textmu m, $r=1$ \textmu m, and $5\leq R_0\leq 100$ \textmu m. The quantity $l(t)+A$ increases exponentially with  time, as predicted by the model, with a mean doubling time of $3.1$ h and a standard deviation of $0.35$ h, shown in Fig. \ref{fig:panel3}A. Fitting is performed only for the time range when tubes are increasing in length. The hypothesis that tube extension is driven by cell division is further supported by the fact that no dependence was found between the tubes' extension rate and initial droplet size.

\section{Discussion}
The deformations of the oil-water interface are the result of the complex interactions between surface tension and the growth of surface-attached bacteria. Without bacteria, interfacial deformations are energetically unfavorable, and droplets remain spherical. The attachment and growth of bacteria leads to multiple effects that promote deformations.

First, attachment of bacteria to the interface reduces the energy of the system by reducing the oil-water surface area. This effect exists for any colloidal particle attached to a liquid-liquid interface. Bacterial attachment being energetically favorable is consistent with our observation of no detachment of cells from the interface before droplets buckle of the experiments. The energy cost of increasing the area of the oil droplet can be mitigated significantly if the surface is covered by bacteria. The magnitude of the reduction in energy depends on the wettability of the cells, which could not be determined in our experiments. Other oil-degrading bacteria, such as \textit{Pseudomonas putida}, have been shown to reduce the oil-water surface tension by more than half as a result of cell hydrophobicity \cite{Ruhs2014}.

Second, the growth of surface-attached cells produces mechanical forces on the interface that oppose surface tension. This effect is similar to what happens when a droplet or bubble covered in colloidal particles shrinks, leading to buckling of the interface \cite{Datta2010,Poulichet2015,Xu2005}. In these systems, a fluid-to-solid transition (jamming) of the colloidal shell occurs because of an increase in the density of particles on the surface. In our system, a similar increase in particle density results from the growth of bacteria. Additionally, as a form of active matter, growing monolayers of rod-shaped bacteria produce collective stresses as a result of their orientational order, particularly at a bounded interface \cite{You2018,You2021}. After the oil-water interface buckles, continued growth of bacteria causes tubular structures to emerge. 

Third, \emph{A. borkumensis} is known to produce biosurfactants at the oil-water interface \cite{Omarova2019,Yakimov1998}, further facilitating interfacial deformations by reducing surface tension. Further research is needed to construct a model that accurately describes the bacteria's collective behavior, including the fluid-to-solid transition that leads to buckling and tubulation. Such a model would include the competing effects of surface tension and active growth stresses, which can be tuned to give rise to the droplet morphologies observed in our experiments.

\subsection{Tubulation and bio-aggregate expansion}
In our experiments, tubulation is observed around droplets of crude or mineral oil covered in rod-shaped bacterial cells. One key difference between the system we describe and those in other studies is that the activity of cells at the oil-water interface is caused solely by cell growth and division, rather than motility.

Our experimental results on the dynamics of tube growth are consistent with the mechanistic model described in \S \ III D, suggesting that tube extension is driven by an exponentially growing monolayer of cells confined to the interface. Additionally, the average extension rate of individual tubular structures from the interface ($3.1\pm0.4$ h) was found to be roughly the same as the mean expansion rate of the bio-aggregate ($3.2\pm1.1$ h). This result supports the hypothesis that bio-aggregate expansion is driven by tube extension. Thus, fundamentally, the emergent phenomena observed in our experiments are driven by by the division of densely packed, rod-shaped cells at a deformable interface. In our previous work, the doubling time of dilute cells on a flat oil-water interface was measured to have a mean of $1.8\pm0.1$ h \cite{Hickl2022}. The slower doubling times observed here suggest that cells at the interface divide more slowly when they are densely packed (such as during tube formation) than when they are dilute. 

For many droplets, the expansion rates measured at different points on the droplet's surface differ by several hours (Fig. \ref{fig:panel3}C). This observation is consistent with the variation in $\Delta R$ around individual droplets (error bars in Fig.~\ref{fig:panel4}B) and can be explained by differences in the direction of tube extension. Images of tubes of oil show that tubes are highly non-linear such as in Fig.~\ref{fig:panel2}. Measured values of both the aggregate expansion rate and its size $\Delta R$ are larger where tubes extend directly radially outward, compared to where they extend from a droplet's surface at an angle. Thus, variation in the experimental values of these variables is consistent with the relatively uniform values of tube extension rate ($3.1\pm0.35$ h) and the cell doubling time ($1.8\pm0.1$ h) observed.

These results show that a growing colony of rod-shaped bacteria confined to a liquid interface can produce morphologies similar to those observed in other active nematic systems. Tube-like protrusions like the ones described here are often associated with topological defects in the orientational order of an active nematic, both in general models \cite{Hoffmann2022} and in specific realizations such as deformable vesicles \cite{Kumar2019} or regenerating epithelial tissues \cite{Maroudas-Sacks2020}. To our knowledge, we are the first to show that tubulation can occur as a result of growing bacterial colonies. Notably, the tubes described here can remain stable for many hours after they form, unlike tubes around deformable vesicles which tend to disappear and retract quickly \cite{Keber2014,Metselaar2019} because of surface tension. Additionally, tube-like protrusions are commonly associated with $+1$ topological defects, whereas the defects in 2D bacterial colonies typically have charges of $\pm1/2$. More work is required to determine how the orientational order of the cells is related to the deformations we observe. By rigorously quantifying both the shape and dynamics of these deformations, we hope to pave the way towards more precise quantitative comparisons between models and experiments of active nematic systems.

\subsection{Implications for oil spill research}

The deformations observed in these experiments could affect the fate of marine spilled oil by increasing the bio-availability of oil to bacteria and influencing the transport of oil droplets. The droplet radii studied here ($5$ \textmu m $\leq R_0\leq 100$ \textmu m) fall in the range observed in deep-sea plumes of oil following the Deepwater Horizon spill in 2010 (approximately $3$ \textmu m $\leq R_0\leq 160$ \textmu m) \cite{Li2015}. Such plumes of oil have been studied\textit{ in situ} \cite{Camilli2010} and in computational models of oil transport \cite{North2015,Paris2012,Lindo-Atichati2016}, with mean droplet radii typically given on the order of 10 \textmu m. However, to our knowledge, models of oil transport have not considered the  interfacial deformations or the expansion of a bio-aggregate that can result from the colonization of the interface by bacteria. Models that attempt to account for oil biodegradation simply treat droplets as shrinking spheres \cite{North2015}. 

Our experiments show that the growth of bacteria on the droplet surface greatly increases the area of the oil-water interface, thereby increasing the bioavailability of oil to bacteria. Based on the average dimensions and density of tubes measured in experiments, the total area of the interface increases by a factor of 6, on average. Additionally, the formation of a bio-aggregate can increase the effective droplet size by as much as a factor of 2. Such expansion would significantly impact the droplet's buoyancy and increase the drag force on it \cite{White2020}. Thus, the effects of bacterial growth must be taken into account to accurately assess the fate of spilled oil \cite{White2019}. The formation of a dense bio-aggregate was observed only for droplets with $R_0\gtrsim15$ \textmu m, demonstrating that droplet size is crucial in determining how the oil is colonized and deformed. Further research is currently being done to assess these effects precisely, both through models of oil transport and laboratory experiments of rising droplets.

Oil can be further dispersed within tubes as it breaks up and coalesces into smaller droplets of average radius $r=1.48$ \textmu m --- more than an order of magnitude smaller than the mean initial droplet radius $R_0$. Mechanically, the breakup of tubes is likely the result of a Plateau-Rayleigh or pearling instability \cite{Bar-Ziv1999,Nelson1995,Alstrom1999}, which causes periodic perturbations in the tube radius to grow over time. In general, surface tension drives droplets to become spherical to minimize the size of the interface, while interfacial cell growth and biosurfactant production drive deformations such as tube formations. After tube extension ends, the oil in the tubes slowly breaks up to form spherical drops embedded in the bioaggregate. This observation suggests that, over time, biosurfactant production by \textit{A. borkumensis} is reduced, and that existing biosurfactant molecules do not remain at the interface indefinitely. Thus, surface tension once again becomes the dominant effect, causing a reduction in the surface area of the oil-water interface. The biological reason for the cessation of biosurfactant production is not know, but could be related to nutrient depletion in the bioaggregate. This phenomenon needs further investigation as it could have important implications for the degradation of these secondary droplets. 

\section{Conclusions}

We conducted experiments to describe the deformations of oil-in-water droplets caused by the active growth of bacteria. As active stresses overcome surface tension, tubes of oil stabilized by bacteria extend exponentially from the interface in order for surface-bound cells to continue dividing. These tubes form a dense bio-aggregate around the droplets, enlarging the droplets significantly. The emergence of tubes and bio-aggregates is highly non-linear, but can be modeled accurately with simple statistical arguments and minimal assumptions. Finally, oil in tubes undergoes breakup into smaller spherical droplets as a result of an instability driven by surface tension. This dispersion of oil significantly increases the surface area of the oil and its bio-availability to the bacteria.

This system is a novel example of active matter physics in a biological system, displaying emergent phenomena that arise from the collective behaviors of a large number of cells far from thermodynamic equilibrium. Our analysis demonstrates how direct visualization combined with models of the system's basic properties can shed light on the underlying physics of complex systems. This research also improves our understanding of the biophysical mechanisms underlying bacterial biodegradation of spilled oil. To our knowledge, the precise deformations of oil droplets due to growth of oil-degrading bacteria had not been described precisely to date. These phenomena are believed to affect both oil degradation rates and transport of droplets through the ocean. Thus, this work can be a stepping stone towards significant improvements in our understanding of the fate of spilled oil in the open ocean.

\section{Conflicts of Interest}

There are no conflicts of interest to declare.

\section{Acknowledgments}

This material is based on work supported by the Oil Spill Recovery Institute under Contract No. 20-10-06. 
We thank Seppe Kuehn and Karna Gowda for useful conversations on cell culturing techniques and Mark Levenstein for useful conversations on data analysis and presentation.

\bibliography{Tubulation_refs}

\clearpage
\newpage


\clearpage
\newpage

\clearpage
\newpage

\clearpage
\newpage

\clearpage
\newpage

\end{document}